\documentclass[a4paper,10pt]{article}
\usepackage{amsmath,amsthm,amssymb}
\usepackage{graphicx,subfigure}
\usepackage{url}

\usepackage{algorithm2e}
\usepackage{listings}
\usepackage{color}

\usepackage{multicol}
\def\twoplot[#1]#2#3#4#5{
\begin{figure}[h]
\begin{multicols}{2}
\begin{center}
    \includegraphics*[#1]{#2}
    \caption{\label{#2} #4}
\end{center}
\begin{center}
    \includegraphics*[#1]{#3}
    \caption{\label{#3} #5}
\end{center}
\end{multicols}
\end{figure}
}

\makeatletter
\def\@citex[#1]#2{\immediate\write\@auxout{\string\citation{#2}}
\def\@citea{}\@cite{\@for\@citeb:=#2\do
{\@citea\def\@citea{; }\@ifundefined
{b@\@citeb}{{\bf ?}\@warning
{Citation `\@citeb' on page \thepage \space undefined}}%
{\csname b@\@citeb\endcsname}}}{#1}}
\makeatother

\usepackage[colorlinks,bookmarksopen,bookmarksnumbered,citecolor=red,urlcolor=red]{hyperref}
\hypersetup{pdftitle={Global sensitivity analysis with 2D hydraulic codes: Applied protocol and practical tool},
bookmarks=true,
pdftoolbar=true,
pdfmenubar=true,
pdfauthor={M. Abily, N. Bertrand, O. Delestre, P. Gourbesville, Y. Richet, C.-M. Duluc},
pdfsubject={Uncertainty, flood hazard modelling, global sensitivity analysis, 2D shallow water equations, Sobol index},
pdfcreator={Delestre},
pdfproducer={Delestre},
pdfkeywords={uncertainty} {flood hazard} {Shallow-Water equation} {Saint-Venant} {global sensitivity analysis} {Sobol index}
{high resolution modelling} {DSM} {surface model} {photogrammetry} {Sobol index}
 {FullSWOF} {Var river} {finite volume} {well-balanced} {hydrostatic reconstruction}}

\title{Global sensitivity analysis with 2D hydraulic codes: Applied protocol and practical tool}

\author{{M. Abily}\footnote{Polytech'Nice Sophia \& URE Innovative-CiTy, University of Nice Sophia Antipolis,
France, e-mail : abily@polytech.unice.fr}, {N. Bertrand}\footnote{Institut de Radioprotection et de S\^uret\'e Nucl\'eaire
(IRSN), PRP-DGE, SCAN, BEHRIG, France, e-mail : nathalie.bertrand@irsn.fr},
{O. Delestre}\footnote{Lab. J.A. Dieudonn\'e UMR7351 CNRS \& EPU Nice Sophia, University of Nice, France, e-mail : delestre@math.unice.fr},
 P. Gourbesville\footnote{Polytech'Nice Sophia \& URE Innovative-CiTy, University of Nice Sophia Antipolis, France},\\ 
 {Y. Richet}\footnote{Institut de Radioprotection et de Sûreté Nucléaire (IRSN), PSN-EXP, SNC, France}
\; and {C.-M. Duluc}\footnote{Institut de Radioprotection et de S\^uret\'e Nucl\'eaire
(IRSN), PRP-DGE, SCAN, BEHRIG}}

\begin{document}
\maketitle

\begin{abstract}
Global Sensitivity Analysis (GSA) methods are useful tools to rank input parameters uncertainties regarding their impact
on result variability. In practice, such type of approach is still at an exploratory level for studies relying on 2D Shallow
Water Equations (SWE) codes as GSA requires specific tools and deals with important computational capacity. The aim of this
paper is to provide both a protocol and a tool to carry out a GSA for 2D hydraulic modelling applications. A coupled tool
between Prom\'eth\'ee (a parametric computation environment) and FullSWOF\_2D (a code relying on 2D SWE) has been set up:
Prom\'eth\'ee-FullSWOF\_2D (P-FS). The main steps of our protocol are: $i$) to identify the 2D hydraulic code input parameters of
interest and to assign them a probability density function, $ii$) to propagate uncertainties within the model, and $iii$) to rank
the effects of each input parameter on the output of interest. For our study case, simulations of a river flood event were
run with uncertainties introduced through three parameters using P-FS tool. Tests were performed on regular computational
mesh, spatially discretizing an urban area, using up to $17.9$ million of computational points. P-FS tool has been installed
on a cluster for computation. Method and P-FS tool successfully allow the computation of Sobol’ indices maps.
\newline
{\bf Keywords} Uncertainty, flood hazard modelling, global sensitivity analysis, 2D shallow water equation, Sobol index.
\newline
\newline
{\bf Analyse globale de sensibilit\'e en mod\'elisation hydraulique \`a surface libre 2D : application d'un protocole
et d\'eveloppement d'outils op\'erationnels --}
Les m\'ethodes d'analyse de sensibilit\'e permettent de contr\^oler la robustesse des r\'esultats de mod\'elisation
ainsi que d'identifier le degr\'e d'influence des param\`etres d'entr\'ee sur le r\'esultat en sortie d'un mod\`ele.
Le processus complet constitue une analyse globale de sensibilit\'e (GSA). Ce type d'approche pr\'esente un grand
int\'er\^et pour analyser les incertitudes de r\'esultats de mod\'elisation, mais est toujours \`a un stade exploratoire
dans les \'etudes appliqu\'ees mettant en jeu des codes bas\'es sur la r\'esolution bidimensionnelle des \'equations
de Saint-Venant. En effet, l'impl\'ementation d'une GSA est d\'elicate car elle n\'ecessite des outils de param\'etrage
automatisable sp\'ecifique et requiers d'importante capacit\'e de calcul. L'objectif de cet article est de pr\'esenter
un protocole et des outils permettant la mise en \oe{}uvre d'une GSA dans des applications de mod\'elisation hydraulique
2D. Un environnement param\'etrique de calcul (Prom\'eth\'ee) et un code de calcul bas\'e sur les \'equations de
Saint-Venant 2D (FullSWOF\_2D) ont \'et\'e adapt\'es pour d\'evelopper l'outil Prom\'eth\'ee-FullSWOF\_2D (P-FS).
Un prototype de protocole op\'erationnel pour la conduite d'une GSA avec un code de calcul d'hydraulique \`a surface
libre 2D est pr\'esent\'e et appliqu\'e \`a un cas test de crue fluvial en milieu urbain. Les \'etapes du protocole
sont : $i$) l'identification des param\`etres hydrauliques 2D d'int\'er\^et et l'attribution d'une loi de probabilit\'e
aux incertitudes associ\'ees \`a ces param\`etres, $ii$) la propagation des incertitudes dans le mod\`ele, et $iii$)
le classement des effets des incertitudes des param\`etres d'entr\'ees sur la variance de la sortie d'int\'er\^et.
Pour le cas test\'e, des simulations d'un sc\'enario ont \'et\'e effectu\'ees avec une incertitude port\'ee sur
trois param\`etres d'entr\'ee. L'outil P-FS a \'et\'e utilis\'e sur un cluster de calcul et est ais\'ement transposable
sur d'autres architectures de calcul intensif. Le protocole de GSA et P-FS ont permis de produire des cartes d'indices
de Sobol afin d'analyser la variabilit\'e spatiale des contributions des param\`etres incertains.
\newline
{\bf Mots cl\'es} Incertitude, mod\'elisation d'inondation extr\^eme, m\'ethode d'analyse de sensibilit\'e, \'equations
de Saint-Venant 2D, indices de Sobol.
\end{abstract}

\section{Introduction}\label{sec:intro}
Sensitivity analysis is an important aspect of the responsible use of hydraulic models. Such approach allows the
identification of the key parameters impacting model performance \cite{Tekatlian01,Iooss11}. Global Sensitivity
Analysis (GSA) aims at ranking the input parameters variability effects on model’s output variability. Only few studies
have performed GSA on hydraulic models. Indeed, this type of approach is not straight forward, as it requires adaptation
of methods and tools development. These aspects being time consuming, not without standing the heavy computational cost of
such type of approach, it represents an important investment for applied practitioners’ community and is therefore still
at an exploratory level.
\newline
Nevertheless, dealing with uncertainties in hydraulic models is a press-forward concern for both practitioners \cite{Iooss11}
and new guidance \cite{ASN13}. In practical flood event modelling applications, mostly uses standard industrial codes relying
on Shallow Water Equations (SWEs) either in 1D ({\it e.g.} Mascaret \cite{Goutal12}, Mike 11 \cite{DHI09} {\it etc.}) or 2D
({\it e.g.} Telemac 2D \cite{Hervouet99}, Mike 21 \cite{DHI07}, Isis2D \cite{Halcrow12} {\it etc.}). The aim of using hydraulic models is
to provide information on simulated flood event properties such as maximal water depth or flood spatial extent. Eventually
outputs of the models are used for design or safety assessment purpose. 
\newline
In hydraulics, deterministic mathematical models aim at representing natural phenomena with different levels of complexity
in the mathematical formulation of the physical phenomena depending on underlying simplifying assumptions. These
assumptions will influence the domain of validity and of application of the models. Consequently, it will impact accuracy
standards which should be expected from models results. An analytical solution to SWE exists from a mathematical point
of view only when the problem is well-posed, which is generally not the case in practical river flood modelling engineering
applications \cite{Cunge14}. Moreover, equations are resolved using computer codes, which will discretely approach the
continuous solutions of these equations (when mathematically existing). Numerical approach implemented in these codes
can be various and different level of accuracy can be expected depending on the numerical method. In deterministic
hydraulic codes, input parameters are variables which are known with a certain level of confidence. Eventually, modeller
choices to design models and computation optimization can introduce high variability in results. In hydraulic models, sources
of uncertainties can be classified in three categories: ($i$) hypothesis in mathematical description of the natural
phenomena, ($ii$) numerical aspects when solving the equations, and ($iii$) input parameters of the model. The uncertainties
related to input parameters are of prime interest for applied practitioners willing to decrease uncertainties in theirs
models results.
\newline
Hydraulic models input parameters are either of hydrological, hydraulic, topographical or numerical type.
Identification, classification and impact quantification of sources of uncertainties, on a given model output, are a set of
analysis steps which will enable to ($i$) analyze uncertainties behavior in a given modelling problem, ($ii$) elaborate
methods for reducing uncertainties on a model output, and ($iii$) communicate on relevant uncertainties. Uncertainty
Analysis (UA) and Sensitivity Analysis (SA) methods are useful tools, as they allow robustness of model predictions to be
checked and help to identify input parameters influences. UA consists in the propagation of uncertainty sources through the
model, and then focus on the quantification of uncertainties in model output and propagating them through the model
predictions \cite{SaintGeours12}. It allows robustness of model results to be checked. Various methods are then available
to rank parameters regarding their impact on results variability (such as Sobol index \cite{Sobol90}). This process goes one
step beyond UA and constitutes a global sensitivity analysis (GSA). In practice, such type of approach is of a great
interest, but is still at an exploratory level in applied studies relying on 2D SWE codes. Indeed, GSA approach
implementation is challenging, as it requires specific tools and deals with important computational capacity. With 1D
free surface hydraulic codes, applied tools and methodology for uncertainty propagation and for GSA have been developed
by IRSN and Companie Nationale du Rh\^one (CNR) \cite{Nguyen15}.
\newline
The purpose of the study presented in this paper is to apply a protocol and to provide a tool, allowing adaptable and
ready-to use GSA for 2D hydraulic modelling applications. Sections developed in this paper present the concept of GSA
applied to 2D hydraulic modelling approach through the presentation ($i$) of the GSA concept, ($ii$) of implemented protocol
and ($iii$) of developed operational tools. A proof of concept to illustrate feasibility of the approach is given, based
on a 2D flood river event modelling in Nice (France) low Var valley.

\section{Material and methods}\label{sec:mat-meth}

\subsection{Global Sensitivity Analysis approach}\label{subsec:GSA-approach}

A regular sensitivity analysis aims to study how the variations of model input parameters impact the models outputs.
Objectives with this approach are mostly to identify the parameter or set of parameters which significantly impact models
outputs \cite{Volkova08,Marrel12}. For instance, SA screening approach is a variance-base method which
allows to identify input variables which have negligible effects, from those which have linear, non-linear or
combinatory effects significantly impacting variability of results output. This can be useful for models with a large
set of input parameters assumed to introduce uncertainty, to reduce the number of input parameters to consider in a
given SA study. Local SA focuses on fixed point in the space of the input. The aim is here to address model behavior
near parameters nominal value to safely assume local linear dependence on the parameter. More details about these SA methods
and their application to practical engineering problems can be found in \cite{Iooss11,Saltelli00,Jacques11}.
\newline
A GSA aims to quantify the output uncertainty in the input factors, given by their uncertainty range and distribution
\cite{Volkova08}. To do so, the deterministic code (2D hydraulic code in our case) is considered as a black box model as described
in \cite{Marrel12}:
\begin{equation}
 \begin{array}{rl}
  f: & \mathbb{R}^p \rightarrow \mathbb{R}\\
    & X \mapsto Y=f(X)\label{eq:BB}
 \end{array}
\end{equation}
where $f$ is the model function, $X = (X_1;...;X_p)$ are $p$ independent input random variables with known distribution
and $Y$ is the output random variable. The principle of GSA method relies on estimation of inputs variables variance
contribution to output variance. A unique functional analysis of variance (ANOVA) decomposition of any integral function
into a sum of elementary functions allows to define the sensitivity indices as explained in
\cite{Sobol90,Marrel12,SaintGeours12}. Sobol’s indices are defined as follow:
\begin{equation}
 S_i=Var\left[ \mathbb{E}(Y|X_i)\right]/Var(Y).\label{eq:Sobol}
\end{equation}
To implement a GSA approach, it is necessary ($i$) to identify inputs and assess their probability distribution, ($ii$) to
propagate uncertainty within the model and ($iii$) to rank the effects of input variability on the output variability
through functional variance decomposition method such as calculation of Sobol indices. The first two steps constitute
an uncertainty analysis (UA) which is a compulsory stage of the GSA.
\newline
For the first step of the UA, each input parameter is considered as a random value where both, choice of considered
input parameter and choice of probability distribution of the input random values, have to be set up. The assessment of
selected uncertain parameters and their probability distribution is completed according to expert opinion or by
statistical analysis for measured values if sufficient measured data sets are available. The first two steps lead to
define probability density function constructed to represent uncertainty of selected input parameters for the study. It
has to be emphasized that this step of the GSA process is important and strongly subjective \cite{Volkova08}.
\newline
Propagation of uncertainty is then required (step $ii$ of the UA), all sources of uncertainties are varied simultaneously,
which is classically done using Monte-Carlo techniques or more parsimonious Monte-Carlo like approach \cite{Iooss11}.
Controlling the convergence of the uncertainty propagation gives an idea if the sample of simulations is large enough
to allow consistent statistical analysis. In practice, convergence of estimated sensitivity indices and their confidence
interval can be plotted and examined visually \cite{Marrel12}. Nevertheless, the decision whether the level of
convergence is satisfactory or not, depends on arbitrary decision of the operator regarding desired accuracy and confidence
interval on the accuracy. Eventually, GSA can be performed to calculate Sobol index \eqref{eq:Sobol}. First-order Sobol index indicates
the contribution to the output variance of the main effect of each input parameters. Total-order Sobol index measures the
contribution to the output variance including all variance caused by interactions between uncertain input parameters
\cite{Iooss11}. The production of Sobol index spatial distribution map is promising. Moreover, such maps have been done in
other application fields such as hydrology, hydrogeology and flood risk cost estimation \cite{SaintGeours12}.

\subsection{Protocol applied for GSA with 2D hydraulic models}\label{subsec:Protocol}

\begin{figure}[htbp]
\begin{center}
\includegraphics[width=0.92\textwidth]{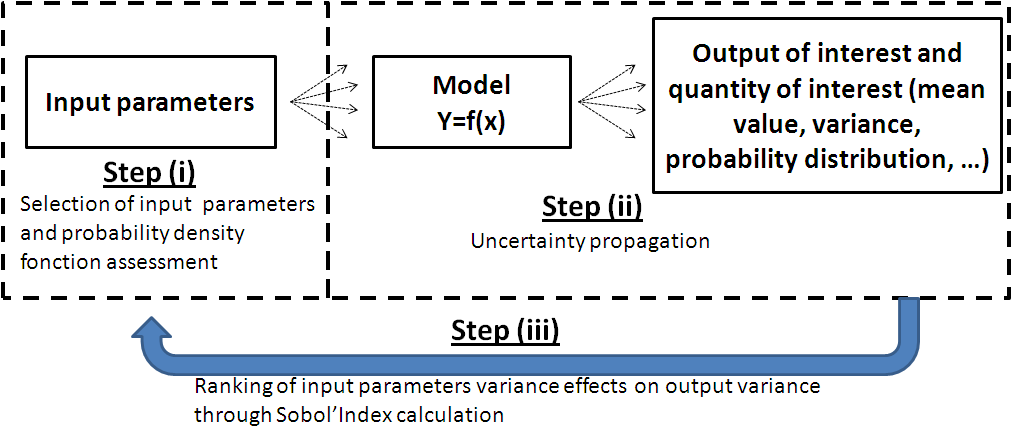}
\caption{Standard GSA approach.}
\label{fig1}
\end{center}
\end{figure}

An overview of GSA approach is presented in fig. \ref{fig1}. This type of general protocol for GSA has already been applied
to 1D hydraulic model \cite{Iooss11,Nguyen15,Alliau15}. Common aspects arise between 1D and 2D
models hydraulic application to implement a GSA approach and are described step by step below:
\newline
The first step (Step $i$, in fig. \ref{fig1}), identifies the input parameters of the hydraulic code to be considered for the
analysis. In hydraulic models, the mains input parameters to consider are: $i$) geometric parameters, spatial discretization,
hydraulic and numerical parameters.
\newline
The geometric parameters, in 1D models, are represented as cross section, whereas in 2D models Digital Elevation Model
(DEM) are included as component of geometric parameter. Another important geometrical aspect to consider as an input
parameter introducing uncertainty is the spatial discretization. The spatial discretization can be considered as a
geometrical and a numerical parameter. Indeed spatial discretization impacts both the level of details of the geometry
included in calculation and numerical stability of calculation as, $dx$ (and $dy$ in 2D) impact CFL criterion. Hydraulic
parameters are mainly initial, boundary conditions and energy loses coefficient. Such parameters are introduced in models
under the form of water level, discharge or flow velocity field or friction coefficients. Geometric and hydraulic variables
can be measured or estimated; as a result they are often subject to a high level of uncertainty.
\newline
Numerical parameters are related to numerical method and solver included in numerical scheme. Broadly speaking, input
numerical parameters will be those related to CFL number ($dt$ and $dx$), parameters impacting accuracy of solver
(numbers of iteration or order of solver method for instance) as well as parameters impacting numerical diffusion
and dispersion.
\newline
Once input parameters are selected, a probability density function has to be attributed to each parameter, in order to
distribute input parameters. As previously mentioned, probability density function are most often related to expert
opinion. The most common used distributions in hydraulic studies are normal or triangular distribution
\cite{Nguyen15}. Uniform distribution is used as well, when no clear idea can be made up regarding the distribution
of a given parameter variability. By assuming equi-probability of the realization $p$ of a variable, uniform
distribution will then maximize the uncertainty (entropy) of the input parameter.
\newline
To apply a GSA, the input parameters are assumed to be independent. This specific point should be considered at this stage
of the GSA, as sometimes selected input parameters can be strongly correlated in some cases. For example, in hydraulic
model dependences between flow and linear energy loses properties (represented by Ch\'ezy, Manning or
Strickler law) are present. 
\newline
The second step (Step $ii$, in fig. \ref{fig1}) results in the propagation of the distributed input parameters within the
selected hydraulic model. Hydraulic codes (1D and 2D) are based on the same sets of equations which are the SWE
\eqref{eq:SV}, as written as follow in 2D \cite{Cunge80}:
\begin{equation}
 \left\{\begin{array}{l}
\partial_t{h}+\partial_x(hu)+\partial_y(hv)=0,\\
\partial_t(hu)+\partial_y(hu^2+gh^2/2)+\partial_y(huv)=gh({S_0}_x-{S_f}_x),\\
\partial_t(hv)+\partial_y(huv)+\partial_y(hv^2+gh^2/2)=gh({S_0}_y-{S_f}_y),
        \end{array}
\right.\label{eq:SV}
\end{equation}
where the unknowns are the velocities $u(x,y,t))$ and $v(x,y,t)$ $\left[\text{m}/\text{s}\right]$ and the water height
$h(x,y,t)$ $\left[m\right]$, and where ${S_0}_x$ (${S_0}_y$) is the opposite of the slope and ${S_f}_x$ (${S_f}_y$)
 the linear energy losses in $x$-direction (resp. $y$-direction).
 \newline
 Therefore, underlying hypothesis in the mathematical description of the physical process are similar for 1D and 2D
 models: ($i$) uniform $u$ (and uniform $v$ in 2D) velocity for a given mesh cell, ($ii$) horizontal free surface at
 a given cell, ($iii$) vertical hydrostatic pressure and (iv) energy losses are represented through Ch\'ezy,
 Manning or Strickler formulas \cite{Cunge80,Chow59}. As previously mentioned, the model is considered
 as a black box as described concerning \eqref{eq:BB} for application of the GSA.
\newline
For each specific source of uncertainty, $n$ independent realizations are generated using Monte-Carlo techniques.
The number of realizations has to be large enough to reach convergence of the interest variable. Histograms are
commonly plotted to ensure that output parameters follow a normal distribution. Moreover, spatially distributed results
of the variable of interest can then be analyzed.
\newline
The third step (Step $iii$, in fig. \ref{fig1}) and final step of GSA approach is the calculation of the Sobol indices and
the evaluation of the model outputs robustness. Analysing GSA results, the model user has a better understanding
and quantification of its models uncertainties.

\subsection{Operational tool and setup}\label{subsec:Operational}

To apply a GSA with 2D Hydraulic models, a coupling between Prom\'eth\'ee a code allowing a parametric environment
of other codes, has been performed with FullSWOF\_2D, a two-dimensional SWE based hydraulic code. The coupling
procedure has taken advantage of previous coupling experience of Prométhée with 1D SWE based hydraulic code
\cite{Nguyen15,Alliau15}. The coupled code Prom\'eth\'ee-FullSWOF (P-FS) has been performed
on a HPC computation structure.

\subsubsection{FullSWOF\_2D}\label{subsubsec:FullSWOF}

FullSWOF\_2D (Full Shallow Water equation for Overland Flow in 2 dimensions) is a code developed as a free software
based on 2D SWE \cite{Delestre14}. In FullSWOF\_2D, the 2D SWE are solved thanks to a well-balanced finite
volume scheme based on the hydrostatic reconstruction \cite{Audusse04c,Delestre10b}. The finite volume
scheme, which is suited for a system of conservation low, is applied on a structured spatial discretization,
using regular Cartesian mesh. For the temporal discretization, a variable time step is used based on the CFL
criterion. The hydrostatic reconstruction (which is a well-balanced numerical strategy) allows to ensure that the
numerical treatment of the system preserves water depth positivity and does not create numerical oscillation
in case of a steady states, where pressures in the flux are balanced with the source term here (topography).
Different solvers can be used HLL, Rusanov, Kinetic \cite{Bouchut04}, VFROE combined with first order or second
order (MUSCL or ENO) reconstruction.
\newline
FullSWOF\_2D is an object oriented software developed in C++. Two parallel versions of the code have been developed
allowing to run calculations under HPC structures \cite{Cordier13}.

\subsubsection{Prom\'eth\'ee-FullSWOF }\label{subsubsec:Promethee-FullSWOF}

Prom\'eth\'ee software is coupled with FullSWOF\_2D. Prom\'eth\'ee is an environment for parametric
computation, allowing carrying out uncertainties propagation study, when coupled to a code. This software is an
open source environment developed by IRSN (\url{http://promethee.irsn.org/doku.php}). The main interest
of Prom\'eth\'ee is the fact that it allows the parameterization of any numerical code. Also, it is optimized
for intensive computing resources use. Moreover, statistical post-treatment can be performed using
Prom\'eth\'ee as it integrates R statistical computing environment \cite{Ihaka98}.
The coupled code Prom\'eth\'ee/FullSWOF (P-FS) is used to automatically launch parameterized computation through R
interface under Linux OS. A graphic user interface is available under Windows OS, but in case of large number of
simulation launching, the use of this OS has shown limitations as described in \cite{Nguyen15}. A maximum of
$30$ calculations can be run simultaneously, with the use of $30$ “daemons”.

\subsection{Practical application test case}\label{subsec:Practical-application}

\begin{figure}[htbp]
\begin{center}
\includegraphics[width=0.98\textwidth]{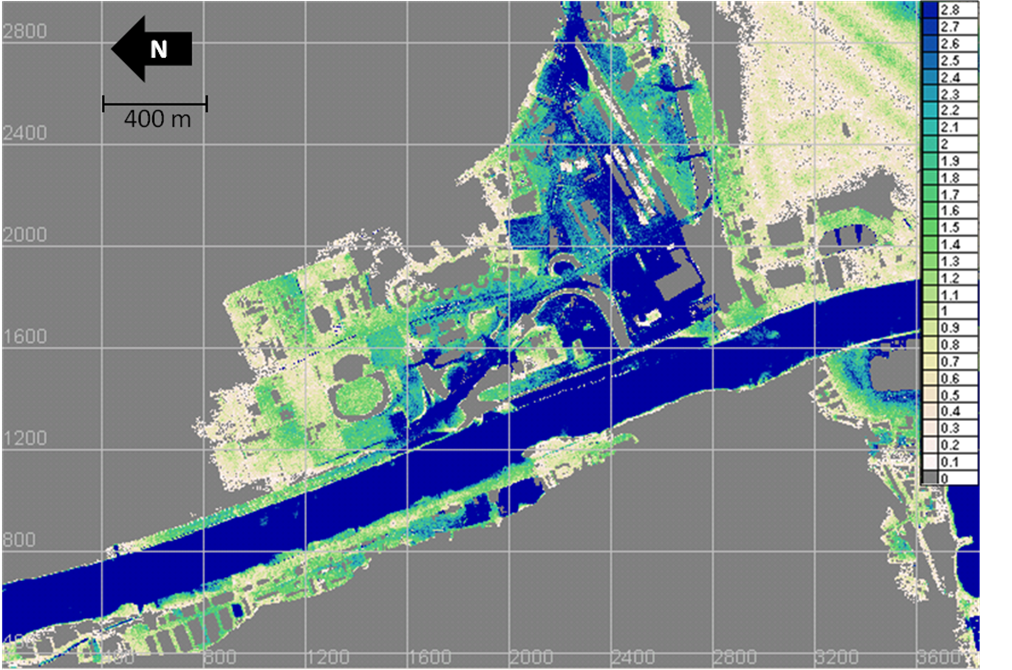}
\caption{Illustration of $h_{max}$ map during an extreme flood event scenario simulated for the low Var river
valley using FullSWOF\_2D.}
\label{fig2}
\end{center}
\end{figure}

To test the GSA protocol and P-FS tools, uncertainties introduced in 2D hydraulic models by geometric input
parameters related to HR topographic data use have been studied. More specifically this case focuses on uncertainty
related to 3D classified data use within hydraulic 2D models. In \cite{Abily16b}, the case study is introduced
and the outcomes of the GSA applied to uncertainties related to high resolution topographic data use with 2D
hydraulic codes are presented in detail. It has to be reminded that the scope of present paper is to give a
proof of concept of possibility to apply protocol and to use developed tools for a GSA with 2D SWE based hydraulic
codes. The main characteristics of that case study are summarized in the current section. 
\newline
The $1994$ flood event of the Var river valley has been modelled. The study area is the lower part of the Var river
valley, where major inundation occurred during the $1994$ event. As our analysis focus on uncertainties related to
geometric input parameters, hydrologic and hydraulic parameters are treated as constant. An illustration of the maximal
water depth ($h_{max}$) computed in this area for the given hydraulic scenario is illustrated in fig. \ref{fig2}. 
\newline
\begin{figure}[htbp]
\begin{center}
\includegraphics[width=0.98\textwidth]{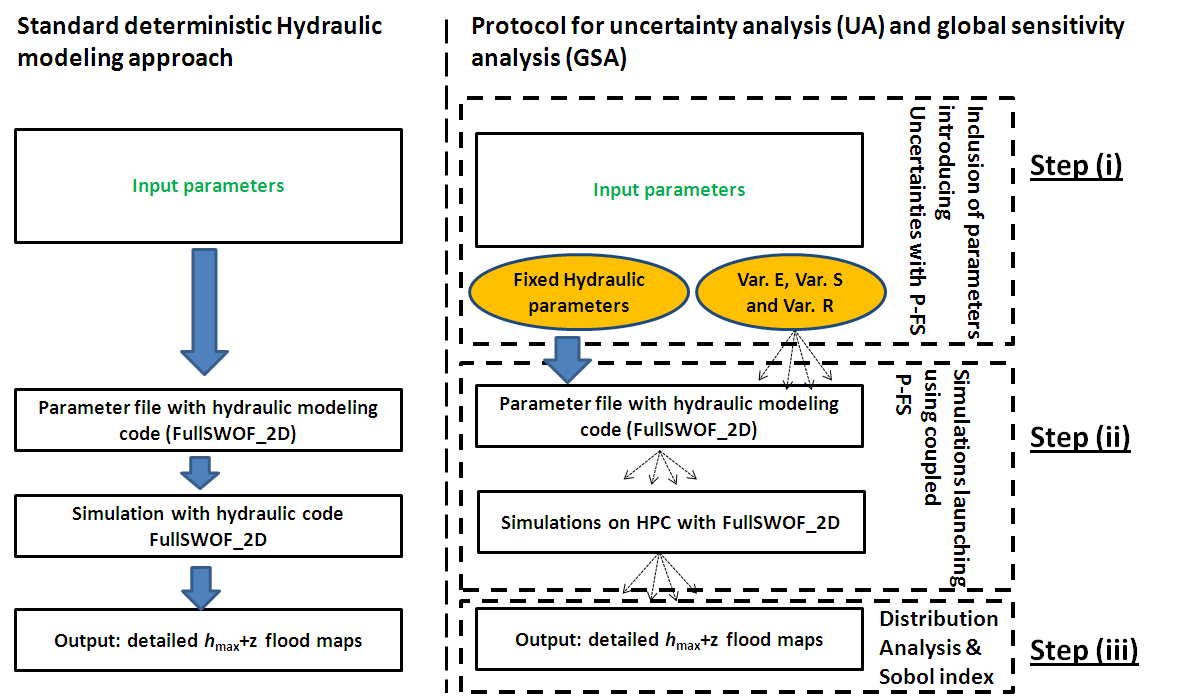}
\caption{Schematization of the GSA protocol adapted to 2D hydraulic modelling.}
\label{fig3}
\end{center}
\end{figure}

Fig. \ref{fig3} illustrates the GSA approach adapted to the tested proof-of-concept 2D hydraulic modelling study. The
three input parameters related to the geometric input parameters are ($i$) the topographic input (called Var. $E$)
and ($ii$) two numerical parameters chosen by modellers (called Var. $S$ and Var. $R$) when willing to use
HR 3D classified data are considered in this GSA practical case.
\newline
Var. $E$ represents measurement errors in topography. Var. $S$ and Var. $R$, are discrete values representing
operator choices, which are respectively concrete elements structures included in DEM (buildings, walls,
street elements {\it etc.}) and structured mesh resolution. These three parameters Var. $E$, $S$ and $R$,
are assumed independent. Var. $E$ is $100$ different occurrences of spatially homogenous $1\;\text{m}$ resolution maps of
errors where each cell of the error grids follows a normal probability density function $\mathcal{N} (0; 0.2)$. A
given occurrence of Var. $E$ is the addition of one of the error grid to a reference High resolution
DTM. Var. $S$ can have $4$ discrete values depending on modeler choices when including HR concrete above ground
elements to DEM. The $S1$ stands for a row DTM, $S2$ for a DEM which encompasses building elevation information,
$S3$ stands for a DEM including $S2$ elevation information plus concrete walls elevations and eventually,
$S4$ is a DEM which including $S3$ information plus Concrete street elements elevation information.
Var. $R$ can have 5 discrete values being $1$, $2$, $3$, $4$ or $5\;\text{m}$, representing spatial discretization choice. 
\newline
As the number of considered parameters is limited, SA screening methods are not considered. The output of
interest is the overland flow water surface elevation ($h_{max}+z$) reached at points and areas of interest.
The GSA ranks influence of selected input parameters variability over variability of $h_{max}+z$.

\section{Results and discussion}\label{sec:results}

\subsection{Feasibility and operational achievement of the approach}\label{subsec:Feasibility-operational}

The introduced GSA approach has been followed with a two-dimensional hydraulic model. The coupled tool has been set
up on the HPC M\'esocentre and the P-FS couple is transposable on any HPC structures. Using R commands,
the simulations were launched. However, the calculation running time of our simulation is very long and increases
considerably when the mesh resolution increases. The computing CPU cost is respectively $2$, $6$, $12$, $24$, $80$ hours
for $5$, $4$, $3$, $2$, $1\;\text{m}$ resolution grid. P-FS tool successfully allow applying GSA protocol to 2D hydraulic modelling.
So far, highest resolution ($R1$ and $R2$) simulations are not fully completed yet, due to a prohibitive
computational time. At these resolutions running simulations on more than $12$ CPU is necessary. Improvement of
parallel version of P-FS is in progress to allow a single simulation to run over more than one node. The R
environment allowed performing post-treatment over the results to analyze them efficiently (see section \ref{subsec:Results-case}).
\newline
The limit of the approach is the computational time required to run the large number of simulations. In our case,
simulations have been run to generate a large set of results, which were sampled afterward, using Monte-Carlo
approach to calculate Sobol indices. Moreover, generation of DEM can be time consuming and cannot be entirely
generated automatically following chosen probability distribution functions and especially when parameters
are discrete. As a result, they have been created before the propagation has been carried out. This remark has
to be extended as a current limit of the tool to generate variable spatial input for 2D hydraulic GSA. This
requires further development. 

\subsection{Results on the illustrative case }\label{subsec:Results-case}

\begin{figure}[htbp]
\begin{center}
\includegraphics[width=0.98\textwidth]{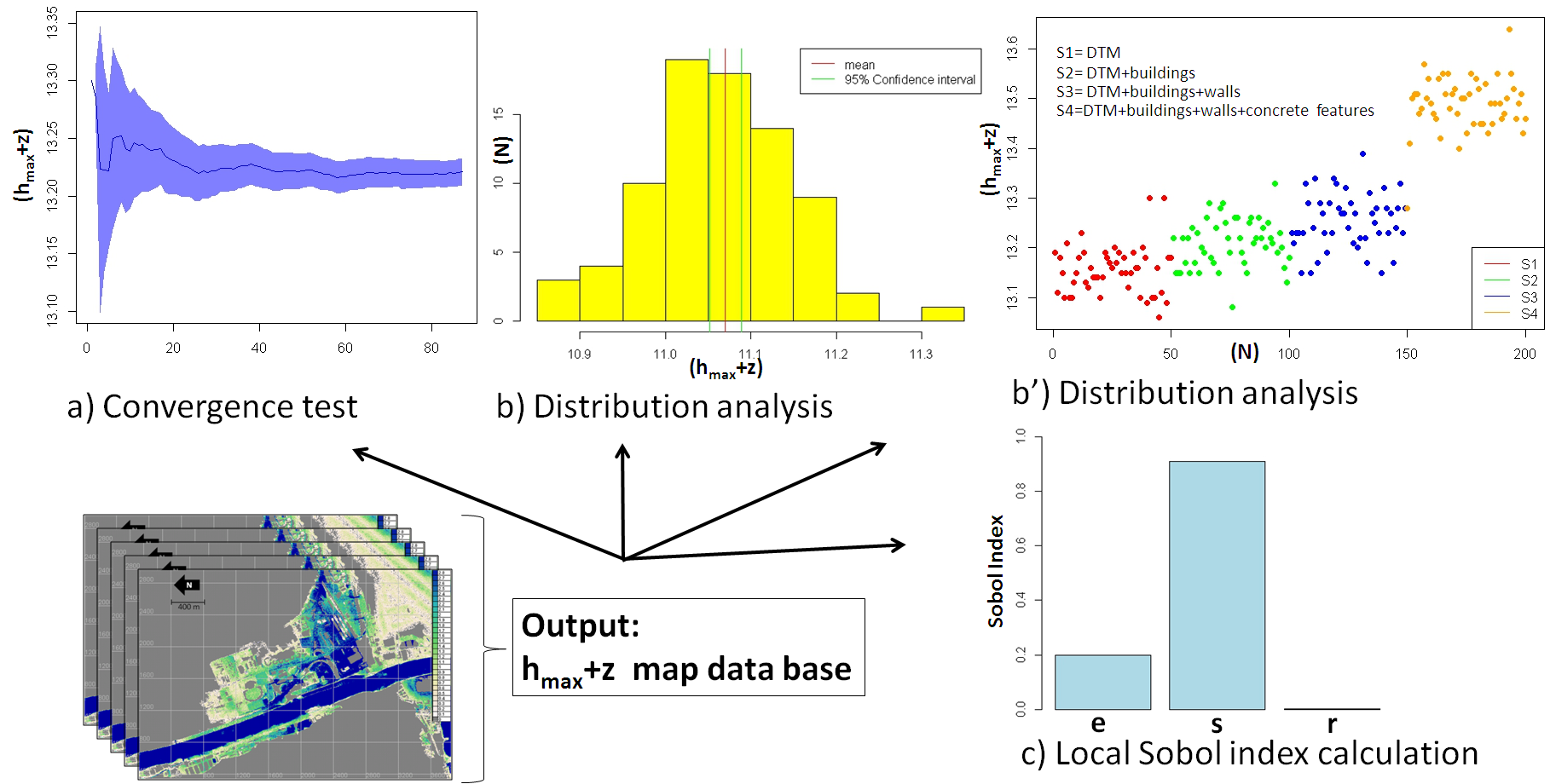}
\caption{Illustration of UA and GSA outputs from the proof-concept study case using protocol and developed P-FS tool.}
\label{fig4}
\end{center}
\end{figure}

The fig. \ref{fig4} introduces the typical results that are obtained when a GSA is performed. Convergence test can be
illustrated by the evolution of mean $h_{max}+z$ value and the $95\%$ confidence interval (CI) when $N$ size
increases (fig. \ref{fig4}.a). Fig. \ref{fig4}.b and \ref{fig4}.b’ respectively illustrate probability density
function analysis of output ($h_{max}+z$) when only one parameter (Var. $E$) is varying (fig. \ref{fig4}.b),
and scatter plotting use for distribution analysis when two parameters (Var. $S$ and $E$) are varying
(fig. \ref{fig4}.b’). First-order Sobol indices are represented in fig. \ref{fig4}.c. Total order Sobol index can
be computed as well. These results have been obtained using R scripts, which were written on purpose for such type
of analysis using existing R functions.
\newline
Nevertheless, it has to be emphasized that limits of the approach are ($i$) needs to fully integrate
specialization possibility in P-FS tool for spatially varying input parameters, ($ii$) computational time
required for such an approach application over resources demanding 2D models can be prohibitive for applied
practitioner and ($iii$) spatialization of output statistical analysis still requires few post treatment
development to allow a fully holistic spatial GSA application for 2D hydraulic models. Eventually an extra
round of analysis and research has to be effectuated using feedback of first results of this approach to
allow improvement of different steps of GSA for 2D hydraulic models (regarding identification of parameters
independence for instance).

\section{Conclusions}\label{sec:conclusions}

In this paper, a GSA framework to investigate the impact of uncertainties of deterministic 2D hydraulic models
input parameters has been developed and tested over the low Var river basin. A coupled tool
Prom\'eth\'ee-FullSWOF\_2D (P-FS) has been elaborated over a standard high performance computation (HPC)
structure. This tool allows parameterization and automatic launching of simulation for uncertainty propagation
step of a GSA. Achieved test on low Var valley constitutes a proof of concept study which has put to the light
the promising possibilities of such an approach for identification of most influent uncertain input parameters.
Indeed, it is possible to go all the way through GSA protocol and after convergence checking, ranking of
influent uncertain input parameter is possible. P-FS is ready to use and easily compatible with most of
HPC structures. Limits and possible improvements of our protocol and tool can be emphasized as follow:
\newline
\begin{enumerate}
 \item Generally speaking, efforts are required for characterization of input parameters spatial variability.
This step of the process can be time consuming and his application in P-FS tool might not be straight forward.
\item Required computational resources to proceed to this type of study are considerable. Here, for the
finest resolutions ($R1$ and $R2$), we had to consider to increase the number of CPUs possible to use
for a given simulation. This will enable to reduce running time of the simulations.
\item The next step is to carry out Sobol index map to illustrate possibilities of protocol and tool use
combination with cross input parameter output variations spatial analysis.
\end{enumerate}

\section*{Acknowledgments}

Photogrametric and photo-interpreted dataset used for this study have been kindly provided by Nice C\^ote d'Azur
Metropolis for research purpose. This work was granted access to (i) the HPC and visualization resources of "Centre
de Calcul Interactif" hosted by "Universit\'e Nice Sophia Antipolis" and (ii) the HPC resources of Aix-Marseille
Universit\'e financed by the project Equip@Meso (ANR-10-EQPX-29-01) of the program "Investissements d'Avenir"
supervised by the Agence Nationale pour la Recherche. Technical support for codes adaptation on high performance
computation centers has been provided by F. Lebas., H. Coullon and P. Navarro.


\end{document}